\documentclass[aps,pra,noshowpacs,reprint,superscriptaddress]{revtex4-1}
\usepackage{graphicx}
\usepackage{bm}
\usepackage{amsmath}

\begin{document}

\title{Ising interaction between two qubits composed of the highest magnetic quantum number states through magnetic dipole-dipole interaction}

\author{Sang Jae Yun}
\email[]{sangjae@kias.re.kr}
\affiliation{School of Computational Sciences, Korea Institute for Advanced Study, Seoul 130-722, Korea}
\affiliation{Center for Relativistic Laser Science, Institute for Basic Science, Gwangju 500-712, Korea}
\author{Jaewan Kim}
\affiliation{School of Computational Sciences, Korea Institute for Advanced Study, Seoul 130-722, Korea}
\author{Chang Hee Nam}
\affiliation{Center for Relativistic Laser Science, Institute for Basic Science, Gwangju 500-712, Korea}
\affiliation{Department of Physics and Photon Science, Gwangju Institute of Science and Technology, Gwangju 500-712, Korea}

\date{\today}

\begin{abstract}
In quantum information processing, one of the most useful interaction between qubits is the Ising type interaction. We propose a scheme to implement the exact Ising interaction through magnetic dipole-dipole interaction. Although magnetic dipolar interaction is Heisenberg type in general, this interaction can bring about the exact mathematical form of the Ising interaction if qubit levels are chosen among the highest magnetic quantum number states. Real physical systems to which our scheme can be applied include rotational states of molecules, hyperfine states of atoms, or electronic states of nitrogen-vacancy centers in diamond. We analyze the feasibility of our scheme for these systems. For example, when the hyperfine levels of rubidium 87 atoms are chosen as qubits and the distance of the two atoms is 0.1 micrometer, controlled-Z gate time will be 8.5 ms. We suggest diverse search and study to achieve optimal implementation of this scheme.
\end{abstract}


\maketitle

\section{Introduction}

In every proposed architecture for quantum information processing, one of the most important issues has been how two-qubit entangling gates are realized \cite{Nielsen2000Book}. If there exists any type of interaction between qubits, in principle the interaction can be used to implement a two-qubit gate \cite{Lloyd1995}. However, for practical purposes, the type of useful interaction is restricted to narrow categories. One of the most preferred is the Ising interaction, which has the form of ${\hat \sigma _{1z}}{\hat \sigma _{2z}}$, because the Ising interaction is mathematically simple so that one can easily design and analyze various quantum gates. To our knowledge, however, no realistic physical system provides mathematically exact form of Ising interaction, but only some effectively approximate forms of Ising interaction have been reported \cite{Mancini2004}. For example, in NMR-based quantum computation, Heisenberg interaction between two nuclear spins is approximated to become Ising interaction by ignoring some extra terms in the magnetic dipole-dipole interaction \cite{Gershenfeld1997}.

The purpose of this paper is to propose a novel scheme that achieves a mathematically exact form of the Ising interaction. We show that the magnetic dipole-dipole interaction reduces to the exact Ising interaction with no approximation if two levels of qubit are chosen among the highest magnetic quantum number states. 

The Ising interaction naturally gives a typical two-qubit gate called controlled-$Z$ gate. Achieving a controlled-$Z$ gate is very useful in the measurement-based quantum computation \cite{Raussendorf2001, Raussendorf2003, Weinstein2005}. In a 2-dimensional arrangement of qubits all prepared in $1/\sqrt 2 \left( {\left|  \downarrow  \right\rangle  + \left|  \uparrow  \right\rangle } \right)$, controlled-$Z$ gate operations between neighboring qubits maximally entangle the entire system into a cluster state, which is the initial state for the measurement-based quantum computation. 

As specific examples to demonstrate our scheme, we present analyses of three physical systems: (i) rotational states of molecular ions trapped in an ion trap, (ii) hyperfine states of atoms trapped in an optical lattice or an ion trap, and (iii) electronic states of nitrogen-vacancy (NV) centers in diamond. The analyses show the following results. The magnetic dipole moment of rotational states of molecules is so small that the interaction time to achieve the controlled-$Z$ gate is too long to be realistic. If the ground hyperfine states of $^{87}{\rm{Rb}}$ atoms are chosen as two levels of qubits, controlled-$Z$ gate time will be 8.5 ms given a distance of 0.1~$\mu {\text{m}}$. Two NV centers separated at 10~nm can give the controlled-Z gate time of $3.6~{\rm{\mu s}}$, but the life time of the qubit state is only about $0.3~{\rm{\mu s}}$. These analyses imply the need of further search for feasible systems.

\section{Choosing qubit levels}

We first choose two highest magnetic quantum number states to constitute a qubit. Angular momentum eigenstates are expressed as $\left| {J,M} \right\rangle $, where $J$ is the angular momentum quantum number and $M$ is its $z$-axis projection in the laboratory frame. $M$ is also called magnetic quantum number. Among $\left| {J,M} \right\rangle $'s, the highest magnetic quantum number states $\left| {J,J} \right\rangle $'s are called directed angular momentum states \cite{Glauber1976, Yun2013Gyro} since their angular momentum directions are well defined along the $z$-axis. We choose qubit levels among the directed angular momentum states such that $\left|  \downarrow  \right\rangle  \equiv \left| {J,J} \right\rangle $ and $\left|  \uparrow  \right\rangle  \equiv \left| {J+n,J+n} \right\rangle $ where $n$ is any nonzero integer such that $J + n \ge 0$. As will be seen later, this particular choice of qubit levels allows us to achieve the {\em exact} Ising interaction through magnetic dipole-dipole interaction.

If a quantum system has angular momentum, the system ordinarily has magnetic dipole moment ${\vec \mu}$ proportional to the angular momentum $\vec J$ with the relation
\begin{equation}
\vec \mu  = \gamma \vec J,
\end{equation}
where ${\gamma}$ is called the gyromagnetic ratio. Even though the system is a composite system composed of several subsystems, Eq.~(1) is still valid if the subsystems are strongly coupled within a small size. Internal individual magnetic moment of each subsystem precesses rapidly (in a semi-classical picture), and the average total magnetic moment appears to be proportional to the total angular momentum. Since a qubit consists of two directed angular momentum states whose angular momentum direction is well defined along the $z$-axis, the direction of the magnetic dipole moment is also well directed along the same direction.

\section{Magnetic dipole-dipole interaction}

\begin{figure}[htb]
        \centering\includegraphics[width=4cm]{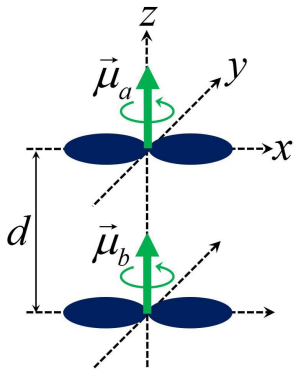}
        \caption{(Color online) Magnetic dipole-dipole interaction in a situation that the centers of two magnetic dipoles lie on the laboratory $z$-axis and their distance is $d$. If qubits consist of two highest magnetc quantum number states, the direction of magnetic dipole moment is well directed along the $z$-axis. }
\end{figure}

We investigate in detail the magnetic dipole-dipole interaction. Two quantum systems with magnetic dipole moments can interact directly with each other. As illustrated in Fig.~1, we consider a situation that the centers of two magnetic dipoles lie on the $z$-axis and their distance is $d$. By using subscripts $a$ and $b$ to indicate the two dipoles, the interaction Hamiltonian for the magnetic dipole-dipole interaction is given by \cite{Fatkullin2010}
\begin{gather}
{{\hat H}_I} = {{{\mu _0}} \over {4\pi {d^3}}}\left[ {{{\hat {\vec \mu} }_a} \cdot {{\hat {\vec \mu} }_b} - 3{{\hat \mu }_{az}}{{\hat \mu }_{bz}}} \right] \nonumber
\\
 =  - {{{\mu _0}{\gamma_a}{\gamma_b}} \over {2\pi {d^3}}}\left[ {{{\hat J}_{az}}{{\hat J}_{bz}} - {1 \over 4}\left( {{{\hat J}_{a + }}{{\hat J}_{b - }} + {{\hat J}_{a - }}{{\hat J}_{b + }}} \right)} \right],
\end{gather}
where ${\mu _0}$ is the vacuum permeability, and ${\hat J_ \pm }$ are the angular momentum ladder operators. In deriving the second line of Eq.~(2), Eq.~(1) is used. Equation~(2) can be simplified further owing to our special choice of the qubit levels. Because our qubit levels consist of only the highest magnetic quantum number states, i.e., $\left| {J,J} \right\rangle $ and $\left| {J+n,J+n} \right\rangle $, the relation ${\hat J_+}|J,J\rangle=0={\hat J_+}|J+n, J+n\rangle$ nullify the second term in the second line of Eq.~(2). This simplifies Eq.~(2) into 
\begin{equation}
{\hat H_I} =  - \eta {\gamma_a}{\gamma_b}{\hat J_{az}}{\hat J_{bz}},
\end{equation}
where $\eta  \equiv {\mu _0}/(2\pi {d^3})$. In other words, Eq.~(2) reduces into an exact Ising interaction by the nullification of raising operators in the Hilbert space made of only the highest magnetic quantum number states. The form of Eq.~(3) is exactly the Ising interaction and is very attractive in quantum information science because it is convenient to implement various quantum gates including not only controlled-$Z$ but also various controlled-phase gates used in quantum Fourier transforms. 

If the qubit levels were chosen not among the highest magnetic quantum number states, the magnetic dipole-dipole interaction would cause transitions between several magnetic quantum number states due to the ladder operators in Eq.~(2). Such transitions would make qubits leak into other magnetic quantum number states outside the qubit Hilbert space and ruin the qubit manipulation. Thus, it is this special choice of the qubit levels that reduces the magnetic dipole-dipole interaction into the exact Ising type.

If the qubits were of pure \emph{spin} angular momentum states where $J$ is fixed, it would not be possible to achieve the Ising interaction. For example, let us consider a $1/2$-spin system whose qubit levels are $\left| {1/2,1/2} \right\rangle $ and $\left| {1/2,-1/2} \right\rangle $. In this case, because raising operator ${\hat J_+}$ cannot kill both states simultaneously, the second term in the second line of Eq.~(2) would not vanish. Thus, Eq.~(3) can arise only when $J$ is plural. Note, however, that the multiplicity does not have to be involved in \emph{orbital} angular momentum states which have infinite number of $J$'s. In other words, the multiplicity can arise in a composite spin system. If one subsystem has spin ${j_1}$ and the other subsystem has ${j_2}$, then the combined spin $j$ can have several values such that $\left| {{j_1} - {j_2}} \right| \le j \le {j_1} + {j_2}$. Whenever one can choose two highest magnetic quantum number states, our scheme works.

\section{Controlled-{\emph Z} gate}

A two-qubit entangling gate can be implemented by the interaction Hamiltonian of Eq.~(3). Since we chose $\left|  \downarrow  \right\rangle  \equiv \left| {J,J} \right\rangle $ and $\left|  \uparrow  \right\rangle  \equiv \left| {J+n,J+n} \right\rangle $, the computational basis of the two-qubit system, $\left\{ {{{\left|  \downarrow  \right\rangle }_a}{{\left|  \downarrow  \right\rangle }_b},{{\left|  \downarrow  \right\rangle }_a}{{\left|  \uparrow  \right\rangle }_b},{{\left|  \uparrow  \right\rangle }_a}{{\left|  \downarrow  \right\rangle }_b},{{\left|  \uparrow  \right\rangle }_a}{{\left|  \uparrow  \right\rangle }_b}} \right\}$, are all eigenstates of Eq.~(3). To investigate eigenvalues, it should be noted that the gyromagnetic ratio $\gamma$ can have different values depending on the qubit levels. This is because, when the qubit is a composite system composed of several subsystems, Land\'e $g$-factor ($g = \gamma \hbar /{\mu _B}$, where ${\mu _B}$ is the Bohr magneton) can have different values depending on the combined angular momentum quantum number. Hence we denote the gyromagnetic ratios of $\left|  \downarrow  \right\rangle $ and $\left|  \uparrow  \right\rangle$ by ${\gamma_\downarrow }$ and ${\gamma_\uparrow }$, respectively. Assuming that the two interacting qubits are the same physical systems (here we are not concerned with the quantum mechanical exchange symmetry of identical particles since the two qubits are well separated in space), the four eigenvalues of the above computational basis are given by ${{\lambda _1} =  - \eta \gamma _ \downarrow ^2{\hbar ^2}{J^2}}$, ${{\lambda _2} = {\lambda _3} =  - \eta {\gamma _ \downarrow }{\gamma _ \uparrow }{\hbar ^2}J(J + n)}$, and ${{\lambda _4} =  - \eta \gamma _ \uparrow ^2{\hbar ^2}{{(J + n)}^2}}$, respectively. 

The matrix representation of the time-evolution operator ${\hat U_I}(t) = \exp ( - i{\hat H}_It/\hbar )$ becomes 
\begin{equation}
\hat U_I(t) = \left( {\begin{matrix}
   {{e^{-i \lambda _1 t/\hbar}}} & 0 & 0 & 0  \cr 
   0 & {{e^{-i \lambda _2 t/\hbar}}} & 0 & 0  \cr 
   0 & 0 & {{e^{-i \lambda _3 t/\hbar}}} & 0  \cr 
   0 & 0 & 0 & {{e^{-i \lambda _4 t/\hbar}}}  \cr 
 \end{matrix}} \right).
\end{equation}
If we denote each diagonal term as $e^{i \phi_k}$ with $k=1,...,4$, $\phi  = {\phi _4} - {\phi _3} - {\phi _2} + {\phi _1}$ is only meaningful because there is a global phase and trivial single particle phases which can be undone by single-qubit operations \cite{Cirac2000}. Simple algebra gives $\phi  = \eta \hbar {\left[ {{\gamma _ \uparrow }(J + n) - {\gamma _ \downarrow }J} \right]^2}t$, and this plays the role of a controlled-phase. At a time satisfying $\phi  = \pi $, Eq.~(4) becomes a controlled-Z operator, a typical two-qubit entangling gate. Hence, the required time for the controlled-Z gate is 
\begin{equation}
{t_{CZ}} = {{2{\pi ^2}{d^3}} \over {{\mu _0}\hbar {{\left[ {{\gamma _ \uparrow }(J + n) - {\gamma _ \downarrow }J} \right]}^2}}}.
\end{equation}
If one can adopt a large $n$, it is advantageous to shorten the gate time, as far as one can perform single-qubit gates between $\left| {J,J} \right\rangle $ and $\left| {J+n,J+n} \right\rangle $. It is noticeable that our entangling gate does not need any laser field during the gate operation. Only approaching the two magnetic dipoles is enough. 

In order to turn off the interaction, it would be enough to separate the two magnetic dipoles. If qubits are made of ions trapped in an ion trap, it is possible to control the distance between the qubits; separation and recombination of two ions has been demonstrated in an array of micro traps \cite{Home2009}. In an optical lattice, distance between the qubits is controllable although not completely separable \cite{Li2008}. Interaction of solid state systems such as color centers in diamond would be impossible to turn off.

\section{Example 1: rotational states of molecular ions}

Now we examine the feasibility of our two-qubit operator scheme by considering a realistic system. We present a detailed analysis for rotational states of molecular ions trapped in an ion trap. A rotational state of a linear molecule is a kind of orbital (spatial) angular momentum states that has multiple angular momentum quantum number. Hence a rotational state of a linear molecule can be a suitable system to apply our scheme. Combined with already established schemes of state preparation, single-qubit operation, and qubit readout in an ion trap setup \cite{Yun2013QC}, our two-qubit operator scheme would satisfy the full set of quantum information processing. Unfortunately, however, our analysis shows \emph {not-so-encouraging} result, which means that magnetic dipole-dipole interaction between two molecular rotors is too weak to perform a two-qubit gate, especially a controlled-$Z$ gate. In spite of this, we present the detailed analysis with the hope that it might give us some guidelines on pursuing other physical systems. 

For rotational states, qubit levels are chosen as $\left|  \downarrow  \right\rangle  \equiv \left| {0,0} \right\rangle $ and $\left|  \uparrow  \right\rangle  \equiv \left| {2,2} \right\rangle $, that is, $\left|  \downarrow  \right\rangle$ is the rotational ground state and $n=2$ is chosen. The reason of choosing $n = 2$ is that single-qubit operations can be easily done by two-photon rotational Raman transition with selection rules $\Delta J =  \pm 2$ and $\Delta M =  \pm 2$ for nonpolar linear molecules \cite{Yun2013Gyro}.

We investigated many kinds of molecules to find a suitable one for our scheme. As mentioned above, we are considering ionic molecules trapped in an ion trap because trapped molecules can be manipulated reliably. Paramagnetic molecules are not suitable because their paramagnetic moment is several thousands times bigger than the rotational magnetic moment so that it causes unwanted magnetic interaction \cite{Yun2013Gyro}. Polar molecules are also excluded since the decoherence time of rotational states of them is only about 0.1 s \cite{Schuster2011}. Among nonparamagnetic nonpolar molecular ions, high-${\gamma}$ (gyromagnetic ratio) molecules are preferred to have large rotational magnetic moments in order to get interactions strong enough. We suggest, as an example, ${\rm{BH}}_{\rm{2}}^{\rm{ + }}$ for a suitable one which has $\gamma_\downarrow  =  \gamma_\uparrow  = - 3.8 \times {10^7}$ ${\rm{rad/(s}} \cdot {\rm{T)}}$ \cite{Dalton2011}. 

Now we estimate the interaction time for controlled-$Z$ gate in a realistic situation using ${\rm{BH}}_{\rm{2}}^{\rm{ + }}$. As expressed in Eq.~(5), the remaining parameter is the distance $d$ between two molecular ions. In ion traps, although typical distance between ions is of order $1~{\rm{ \mu m}}$, it was reported that an ion trap was designed to approach ions to a distance of $d = 0.1~{\rm{ \mu m}}$ \cite{DeVoe1996}. Assuming this distance, ${\rm{BH}}_{\rm{2}}^{\rm{ + }}$ takes ${t_{CZ}} = 26010$ s to achieve the controlled-$Z$ gate. This operation time is too long to be a practical gate. Since we took an extreme distance hard to achieve in an ion trap and chose a high-${\gamma}$ molecule intentionally, we conclude that the rotational states of molecular ions are not suitable for our scheme to be applied.

\section{Example 2: hyperfine states of atoms}

\begin{figure}[tbh]
        \centering\includegraphics[width=6cm]{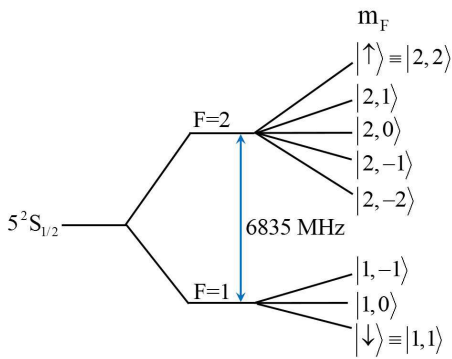}
        \caption{Hyperfine level structure of the electronic ground state ${\rm{5}}{}^{\rm{2}}{{\rm{S}}_{{\rm{1/2}}}}$ of ${}^{87}{\rm{Rb}}$. Qubit levels consist of the two highest magnetic quantum number states.}
\end{figure}

Hyperfine levels of atoms in the electronic ground state have plural angular momentum quantum number $F$ combined by the electronic spin and the nuclear spin. Here we conduct a concrete analysis by choosing a particular atom ${}^{87}{\rm{Rb}}$ trapped in an optical lattice \cite{Duan2003}. As shown in Fig.~2, we choose qubit levels such that $\left|  \downarrow  \right\rangle  \equiv \left| {F = 1,{m_F} = 1} \right\rangle $ and $\left|  \uparrow  \right\rangle  \equiv \left| {F = 2,{m_F} = 2} \right\rangle $. Land\'e $g$-factor of the $F=1$ states is $-0.5$ and that of $F=2$ states is $0.5$, which correspond to ${\gamma _ \downarrow } =  - 4.4 \times {10^{10}}$~${\rm{rad/}}\left( {{\rm{s}} \cdot {\rm{T}}} \right)$ and ${\gamma _ \uparrow } = 4.4 \times {10^{10}}$~${\rm{rad/}}\left( {{\rm{s}} \cdot {\rm{T}}} \right)$, respectively. Note that this value is thousands times bigger than that of the molecular rotors. If the distance $d$ between the two ${}^{87}{\rm{Rb}}$ atoms is $1~{\rm{\mu m}}$, the controlled-Z gate time will be ${t_{CZ}} = 8.5~{\rm{s}}$ by Eq. (5). If $d=0.1~{\rm{\mu m}}$, ${t_{CZ}} = 8.5~{\rm{ms}}$.

The decoherence time of the hyperfine levels of ${}^{87}{\rm{Rb}}$ in an optical lattice reached 21 s \cite{Kleine2011}, but their qubit levels were $\left|  \downarrow  \right\rangle  \equiv \left| {F = 1,{m_F} = 0} \right\rangle $ and $\left|  \uparrow  \right\rangle  \equiv \left| {F = 2,{m_F} = 0} \right\rangle $ different from our choice. Despite such difference, we expect that our qubit levels can also achieve a enough coherence time longer than ${t_{CZ}}$. It needs further study for the decoherence time of our qubit levels. 

Ionic atoms such as ${}^{9}{\rm{B}}{{\rm{e}}^ + }$ also have a good hyperfine level structure \cite{Langer2005} that can be adopted to our scheme. The analysis should be similar to ${}^{87}{\rm{Rb}}$. The only difference is that ion trap is used instead of optical lattice for trapping. In both optical lattices for neutral atoms and ion traps for ions, state preparation, single-qubit operation and qubit readout could also be done by the well-developed techniques manipulating hyperfine levels.

\section{Example 3: nitrogen-vacancy centers in diamond}

\begin{figure}[tbh]
        \centering\includegraphics[width=5cm]{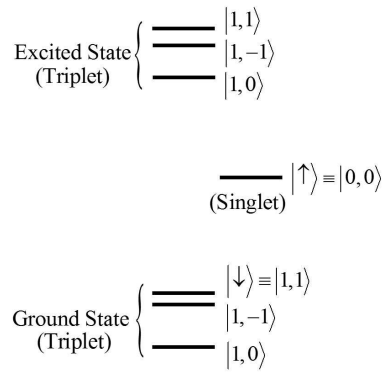}
        \caption{Energy level structure of the nitrogen-vacancy center comprising ground state and excited state triplets and an intermediate singlet state.}
\end{figure}

Now we consider nitrogen-vacancy (NV) centers in diamond. A big advantage of NV centers is that they can reside very closely \cite{Neumann2010, Dolde2013}. $d = 10~{\rm{nm}}$ has been realized experimentally \cite{Neumann2010}. Qubit levels are chosen as shown in Fig.~3. $\left|  \downarrow  \right\rangle  \equiv \left| {1,1} \right\rangle $ is one of the ground triplet state  whose $g$-factor is 2.3 \cite{Neumann2010}, and $\left|  \uparrow  \right\rangle  \equiv \left| {0,0} \right\rangle $ is the intermediate state whose $g$-factor is zero because it is a singlet state. With these values and $d=\rm{10~nm}$, we get ${t_{CZ}} = 3.6~{\rm{\mu s}}$ by Eq. (5). 

Unfortunately, the lifetime of the intermediate singlet state, i.e., $\left|  \uparrow  \right\rangle $, is about 0.3~$\rm{\mu s}$ \cite{Manson2006}. Thus, our choice of qubit levels in NV centers would hardly achieve a high-fidelity two-qubit gate. However, it would be worth investigating other kinds of color centers in diamond whether some of them would be adoptable to our scheme.

\section{Conclusion}

We have shown that, if qubit levels are chosen among the highest magnetic quantum number states of an angular momentum system, magnetic dipole-dipole interaction becomes exact Ising interaction without any approximation. The highest magnetic quantum number states, $\left| {J,J} \right\rangle $ and $\left| {J+n,J+n} \right\rangle $, are not necessarily confined to orbital angular momentum. A combined spin system produced by addition of two spin systems can also be adopted. Thus, our scheme of two-qubit operation could be applied to a variety of physical systems. As realistic examples, we have analyzed the feasibility of our scheme on the rotational states of molecules, hyperfine states of atoms, and electronic states of nitrogen-vacancy centers in diamond. Although we have provided rough estimations, we have gained some guidelines for searching physical systems to which our scheme is applicable. The crucial parameters are the distance ${d}$ and the gyromagnetic ratio $\gamma$ in order to get the Ising interaction strong enough to achieve the controlled-$Z$ gate operation. If the distance between the two magnetic dipoles approaches 10 nm, the controlled-Z gate operation can be achieved in an order of microsecond. It is expected that this study would stimulate more detailed analyses on various physical systems.

\begin{acknowledgments}
We appreciate Jaeyoon Cho for helpful comments. This work was supported by IBS (Institute for Basic Science) under IBS-R012-D1, and was partly supported by the IT R\&D program of MOTIE/KEIT [10043464 (2013)].
\end{acknowledgments}

\bibliography{Bib_total}


\end{document}